\documentclass[12pt,twoside,dvips]{article}
\usepackage{epsfig}
\usepackage{a4wide}
\usepackage{graphics}
\usepackage{epsfig}
\usepackage{rotating}
\usepackage{pifont}
\usepackage{epsf}
\usepackage{psboxit}

\PScommands

\newcommand{\gsim}{\rlap{\raise 2pt \hbox{$>$}}{\lower 2pt \hbox{$\sim$}}}
\newcommand{\cosscat}{\mbox{$\cos\Theta_{\overline{\Lambda}}^{*}$}}
\newcommand{\lam}{\mbox{$\Lambda$}}
\newcommand{\lamb}{\mbox{$\overline{\Lambda}$}}
\newcommand{\pbar}{\mbox{$\overline{p}$}}
\newcommand{\llb}{\mbox{$\overline{p}p\rightarrow\overline{\Lambda}\Lambda$}}

\begin{document}
\mbox{ }
\vspace{20mm}

\includegraphics{tslisvhead.eps}

\begin{flushright}
\begin{minipage}[t]{42mm}
{\bf TSL/ISV-2002-0262 \\
April 2002} 
\end{minipage}
\end{flushright}

\begin{center}
\vspace{10mm}

{\bf \Large String model description of polarisation and
   angular distributions in {\boldmath\llb} 
   at low energies
}

\vspace{15mm}

{\Large S. Pomp$^{\dagger}$, G. Ingelman, 
T. Johansson and S. Ohlsson$^{\ddagger}$ \\ }
\vspace{3mm}
     Dept. of Radiation Sciences, Uppsala University, 
     Box 535, S--75121 Uppsala, Sweden \\
$^\dagger$ now at Dept. of Neutron Research, Uppsala University,
     S--75120 Uppsala, Sweden \\
$^\ddagger$ now at ESRF, F--38043 Grenoble, France \\ 

\end{center}

\vspace{10mm}

{\bf Abstract:}
The observed polarisation of $\Lambda$ hyperons from the inclusive 
$pA \rightarrow \Lambda X$ reaction at high 
energies has previously been well described within the Lund string model 
through polarised $s\bar{s}$ quark pair production in the string breaking 
hadronisation process. This model is here applied to the exclusive {\llb} 
reaction at low energies and compared to available data sets down to an 
incident beam momentum of 1.835 GeV/c. 
This required an extension of the diquark scattering 
model to involve three components: an isotropic part relevant close 
to threshold,
a spectator part and a forward scattering part as 
in $pA \rightarrow \Lambda X$ 
at high energies. The observed angular distributions are then reproduced 
and, for momentum transfers above \mbox{$|t'| = 0.2$ GeV$^{2}$}, 
agreement with the measured polarisation is also obtained. 

\vspace{5mm}

PACS numbers: 13.88.+e, 14.20.Jn


%
%

\vfill
\clearpage
\setcounter{page}{1}

\section{Introduction}
\label{sec:intro}
%

In the mid 70's it was discovered that $\Lambda$ hyperons
produced by an unpolarised high--energy proton beam on
an unpolarised target nucleus are polarised~\cite{lesnik,bunce}.
This came as a big surprise since it was believed that
spin effects would not survive at these high energies
and perturbative QCD was indeed not able to describe
the observations~\cite{kane,barni}.

Today, more than 20 years after these first observations,
similar polarisation effects have been observed
in a large energy range and for other hyperons of the baryon octet
($\Sigma^-$~\cite{deck},
  $\Sigma^0$~\cite{sigma0},
  $\Sigma^+$~\cite{wilkinson},
  $\Xi^-$~\cite{xi-} and
  $\Xi^0$~\cite{xi0}).
The trend of the data can be summarised as
\[
    P(\Lambda)     \approx 
    P(\Xi^-)       \approx 
    P(\Xi^0)       \approx 
   -P(\Sigma^-)    \approx  
   -P(\Sigma^0)    \approx  
   -P(\Sigma^+).
\]
The understanding of the mechanism that produces polarised
hyperons is, however, still far from complete and a wealth of different
models have been proposed \cite{heller77,lund,szwed,degrand,boros2,toki}.
The model discussed in this paper is an extension of
a semiclassical model \cite{lund} that provides a good description of 
observed polarisation data for inclusively produced $\Lambda$ hyperons in
$pA \rightarrow \Lambda X$ reactions. The model is based on the 
Lund string model \cite{stringmodel}, 
which successfully describes the hadronisation of quarks and gluons emerging 
from high energy interactions. 

Most of the experimental data on hyperon polarisation come indeed from inclusive
studies, but more recently also exclusive reactions
have been studied~\cite{ps185,felix,balestra99},
thus removing the uncertainty whether the observed {\lam}'s have been produced
directly or stem from decays of heavier hyperons.
In this paper we investigate whether the string-based model can be extended to 
describe exclusive hyperon production in the {\llb} reaction.
Data obtained by the PS185 collaboration at LEAR, CERN,
for this reaction~\cite{ps185} are of high precision and allow the validity of 
the model to be tested with high accuracy in the low energy domain
(beam momenta below 2.0~GeV/c).

\section{Basics of the string model}
\label{sec:model}

\begin{figure}[htp!]
 \begin{center}
 \resizebox{0.75\textwidth}{!}
 {
 \begin{picture}(280,160)(0,-50)
  \put(10,130){\makebox(0,0){a)}}
  \thinlines
  \thicklines
  \put(0,20){\vector(1,0){30}} 
  \put(10,28){\makebox(0,0){$\vec{p}_{p}$}}
  \put(70,20){\circle*{10}}
  \put(70,20){\line(2,1){60}} 
  \put(130,50){\circle*{4}}
  \put(170,70){\circle*{4}}
  \put(170,70){\line(2,1){40}} 
  \put(209,92){\circle*{4}}
  \put(211,88){\circle*{4}}
  \thinlines
  \put(70,20){\vector(2,1){200}} 
  \put(270,90){\vector(0,1){30}} 
  \put(150,60){\circle*{2}}
  \put(150,60){\circle{6}}
  \put(130,50){\vector(1,-2){12}} 
  \put(170,70){\vector(-1,2){12}} 
  \multiput(33,20)(10,0){25}{\line(1,0){5}}
  \multiput(218,90)(10,0){7}{\line(1,0){5}}
  \put(177,67){\makebox(0,0){$s$}}
  \put(123,53){\makebox(0,0){$\overline{s}$}}
  \put(171,94){\makebox(0,0){$\vec{k}_{T}$}}
  \put(153,31){\makebox(0,0){$-\vec{k}_{T}$}}
  \put(143,69){\makebox(0,0){$\vec{L}$}}
  \put(210,104){\makebox(0,0){$(ud)_{0}$}}
  \put(258,123){\makebox(0,0){$\vec{q}$}}
  \put(279,110){\makebox(0,0){$\vec{q}_{T}$}}
 \end{picture}
 }
%
%
 \resizebox{0.75\textwidth}{!}
 {
 \begin{picture}(280,180)(0,-20)
  \put(10,125){\makebox(0,0){b)}}
  \thinlines
  \thicklines
  \put(0,60){\vector(1,0){30}}
  \put(10,68){\makebox(0,0){$\vec{p}_{\overline{p}}$}}
  \put(280,60){\vector(-1,0){30}}
  \put(270,68){\makebox(0,0){$\vec{p}_{p}$}}
  \put(90,30){\line(2,1){40}}
  \put(130,50){\circle*{4}}
  \put(170,70){\circle*{4}}
  \put(170,70){\line(2,1){40}}
  \put(89,32){\circle*{4}}
  \put(91,28){\circle*{4}}
  \put(150,60){\circle*{2}}
  \put(150,60){\circle{6}}
  \put(209,92){\circle*{4}}
  \put(211,88){\circle*{4}}
  \thinlines
  \put(150,60){\vector(2,1){120}}
  \put(150,60){\vector(-2,-1){120}}
  \put(270,90){\vector(0,1){30}}
  \put(130,50){\vector(1,-2){12}}
  \put(170,70){\vector(-1,2){12}}
  \multiput(33,60)(10,0){25}{\line(1,0){5}}
  \multiput(218,90)(10,0){7}{\line(1,0){5}}
  \put(177,67){\makebox(0,0){$\overline{s}$}}
  \put(123,53){\makebox(0,0){$s$}}
  \put(171,94){\makebox(0,0){$\vec{k}_{T}$}}
  \put(153,31){\makebox(0,0){$-\vec{k}_{T}$}}
  \put(143,69){\makebox(0,0){$\vec{L}$}}
  \put(70,35){\makebox(0,0){$(ud)_{0}$}}
  \put(210,104){\makebox(0,0){$(\overline{u}\overline{d})_{0}$}}
  \put(258,123){\makebox(0,0){$\vec{q}$}}
  \put(40,20){\makebox(0,0){$-\vec{q}$}}
  \put(279,110){\makebox(0,0){$\vec{q}_{T}$}}
 \end{picture}
 }
 \end{center}
 \caption{
   (a) Inclusive $\Lambda$ production in the string model: 
   A forward-scattered $(ud)_0$ 
   diquark stretches a string field producing an $s\bar{s}$ pair, whose  
   transverse momenta introduce an orbital angular momentum $\vec{L}$ 
   pointing out of the scattering plane. 
   To conserve the zero angular momentum in the string field,
   the $s$ and $\bar{s}$ spins are polarised in opposite direction.  
   \protect\newline
   (b) The string model adapted to {\llb}: A $u$ and $\bar{u}$ annihilates 
   giving $ud$ and $\bar{u}\bar{d}$ diquarks moving apart with opposite momenta
   ($q$) and an $s\bar{s}$ pair is produced (as in a), but with kinematical 
   constraints for the exclusive production of $\Lambda \bar{\Lambda}$ in 
   the c.m. system. 
 }
 \label{fig:lund_mod2}
\end{figure}
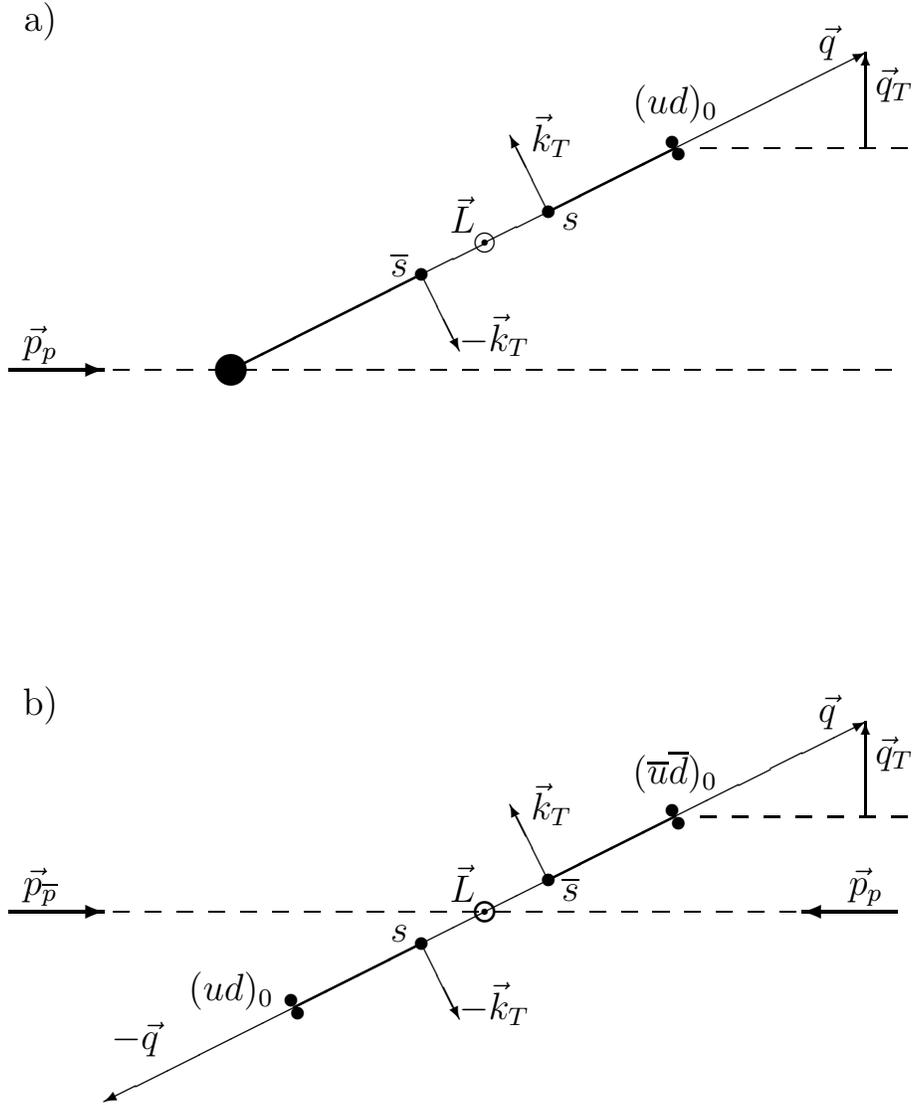

The original string model for polarisation is illustrated in 
Fig.~\ref{fig:lund_mod2}a and described in this section.

 The incoming proton interacts with a target nucleon resulting in a 
  forward-moving diquark system with transverse momentum $q_{T}$.
  A colour triplet string-field, with no transverse degrees of
  freedom, is stretched from this diquark
  (having an antitriplet colour charge) to a colour triplet charge
  in the collision region.

 The potential energy in the field is reduced by breaking the 
  string through the production of quark--antiquark pairs, {\it i.e.},
  colour triplet-antitriplet charges. 
  Within the SU(6) constituent quark model, a {\lam} consists of a
  $(ud)_{0}$ diquark ($I = S = 0$) and an $s$ quark.
  Therefore, if the scattered diquark is a $(ud)_{0}$ quark pair
  and a strange--antistrange quark pair is produced in the string breaking,
  then a {\lam} can be produced.

 Transverse momentum is locally conserved 
  and the $s$ and $\bar{s}$ quarks are therefore 
  produced with equal but oppositely directed transverse
  momenta, $k_{T}$, relative to the string.
  For small scattering angles, the transverse momentum of the {\lam} becomes
  $\vec{p}_{T} \approx \vec{q}_{T} + \vec{k}_{T}.$
  The distributions in $k_{T}$ and $q_{T}$ are taken to be gaussian 
  having widths $\sigma_{q_T}$ and $\sigma_{k_T}$, respectively. 
  To account for $k_{T}$ and the strange quark mass
  \mbox{$m_{s} \approx 0.2$ GeV/c$^2$},
  the $s$ and $\bar{s}$ are produced a distance 
  \mbox{$l = 2 \sqrt{m_{s}^{2} + k_{T}^{2}} / \kappa$} 
  apart, such that energy stored in the string 
  ($\kappa \approx 1 \: GeV/fm$) 
  is transformed to the transverse mass of the produced pair. 
  This can be described as a virtual $s\bar{s}$ 
  pair being produced in a single space--time point and then tunneling 
  apart through the potential until they become on--shell. 
  Treating this quantum mechanical tunneling process with the WKB method
  gives the gaussian $k_T$ spectrum used. 

 The separated $\vec{k}_{T}$--vectors produce an orbital angular momentum 
  $L \:=\: lk_{T}$ of the $s\bar{s}$ pair. The key assumption of the 
  model is that this orbital angular momentum is compensated by 
  the polarisation of the spin of the $s$ and $\bar{s}$ in order to 
  locally conserve the total angular momentum, which was zero in the string
  without transverse degrees of freedom. 
  Since it is the $s$ quark that carries the spin of the {\lam}, the end result 
  is the production of polarised {\lam}'s. 

The model's correlation between $k_{T}$ and the spin polarisation of the $s$ 
quark results in a {\lam} polarisation which increase with its transverse 
momentum $p_T$. 
This may be viewed as a trigger bias effect, where a sample of {\lam}'s with 
a certain value of $\vec{p}_{T}$ will have an enhanced number of events where 
$\vec{k}_{T}$ points in the same direction and hence having a net polarisation. 

In order for the compensation of orbital angular momenta by the quark spins to 
be consistent, it is necessary that the $k_{T}$ distribution is such that 
$L>1$ are suppressed. This is indeed the case, since $L > 1$ corresponds to 
$k_{T} \gsim 0.3$ GeV/c with the values for $m_{s}$ and $\kappa$ quoted above. 
The dependence of the polarisation $P_{q}$ on $L$ is parameterised as
$ P_{q} \: = \: L /(\beta + L)$, with $\beta \sim 1$, such that the polarisation
increases linearly for small $L$ and saturates at 100\%  for large $L$. 
The results are, however, rather insensitive to the particular function chosen 
to describe the $P_{q}$--$L$ correlation \cite{lund}.
The model is numerically evaluated through a Monte Carlo simulation 
specifying momenta and polarisation event by event. 

\section{Model simulations for {\lamb\lam} production}
\label{sec:lund-data}

Measurements of the exclusive reaction {\llb} gives additional information
as compared to inclusive $\Lambda$ production. 
The two-body kinematics imposes strong constraints that fixes the 
absolute value 
of the momenta of the two hyperons in the c.m.~system. 
In order to apply the string--based model described above, 
it must be extended as illustrated in Fig.~\ref{fig:lund_mod2}b. 
In the $p\bar{p}$ interaction a $u$ and $\bar{u}$ 
are annihilated leaving the $ud$ and $\bar{u}\bar{d}$ moving in opposite 
directions with equal momenta. 
Due to the fixed $\overline{\Lambda}$ ($\Lambda$)
c.m.~momentum, the diquark momentum $\vec{q}$ ($-\vec{q}$)
is no longer independent of the momentum $\vec{k}_{T}$ ($-\vec{k}_{T}$)
of the $\bar{s}$ ($s$) quark. 

An important point of the above polarisation mechanism is that the $s$ and 
$\bar{s}$ have parallel spins and are therefore in a spin triplet state. 
Consequently, the model gives a natural explanation for 
the experimental fact~\cite{ps185_leap} that 
the $\overline{\Lambda}\Lambda$ are dominantly in a spin triplet state. 

\begin{figure}[bthp!]
  \begin{center}
  \resizebox{0.85\textwidth}{!}
  {\includegraphics*[1cm,5cm][20cm,21.5cm]{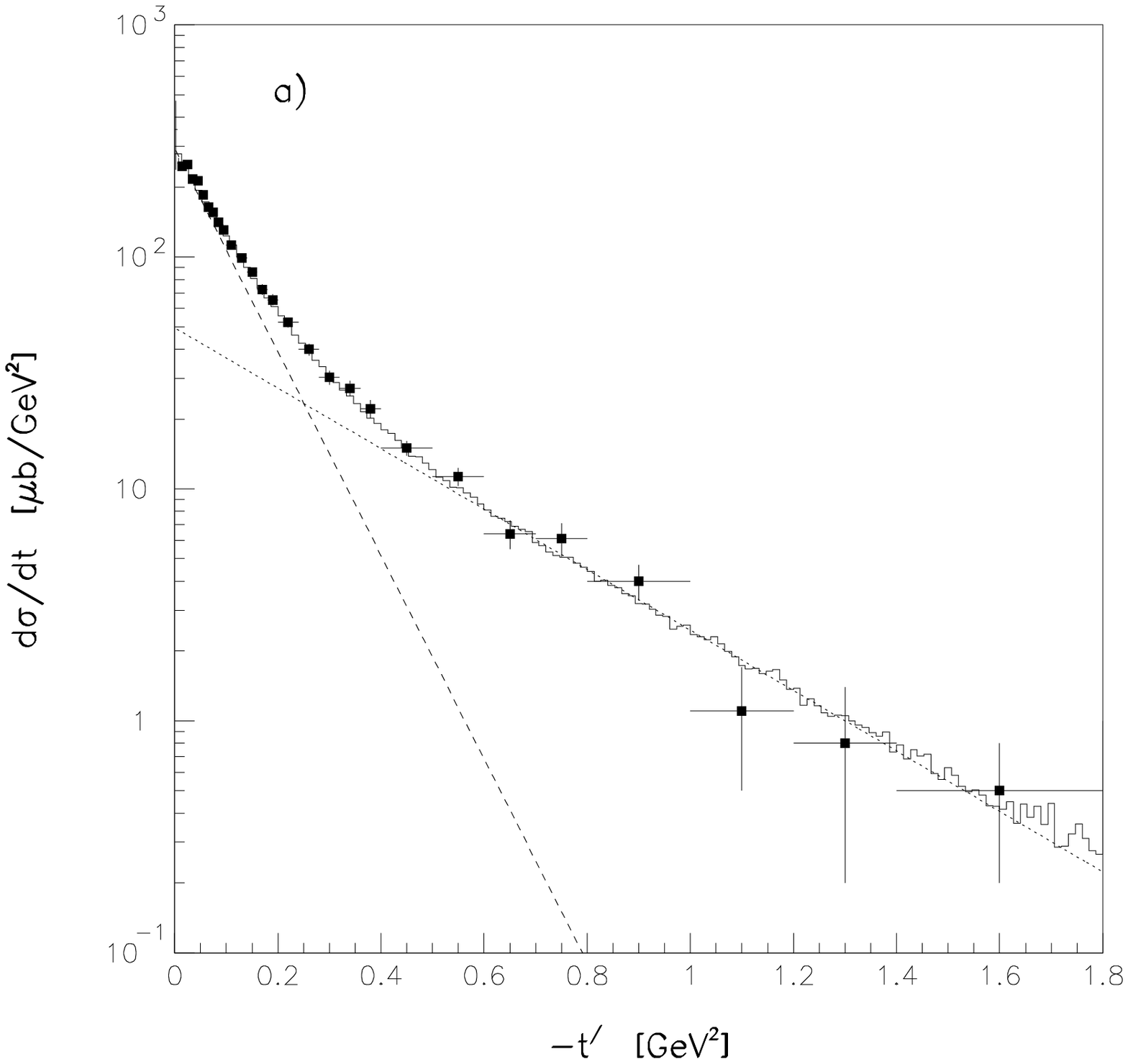}}
  \resizebox{0.85\textwidth}{!}
  {\includegraphics*[1cm,13cm][20cm,22cm]{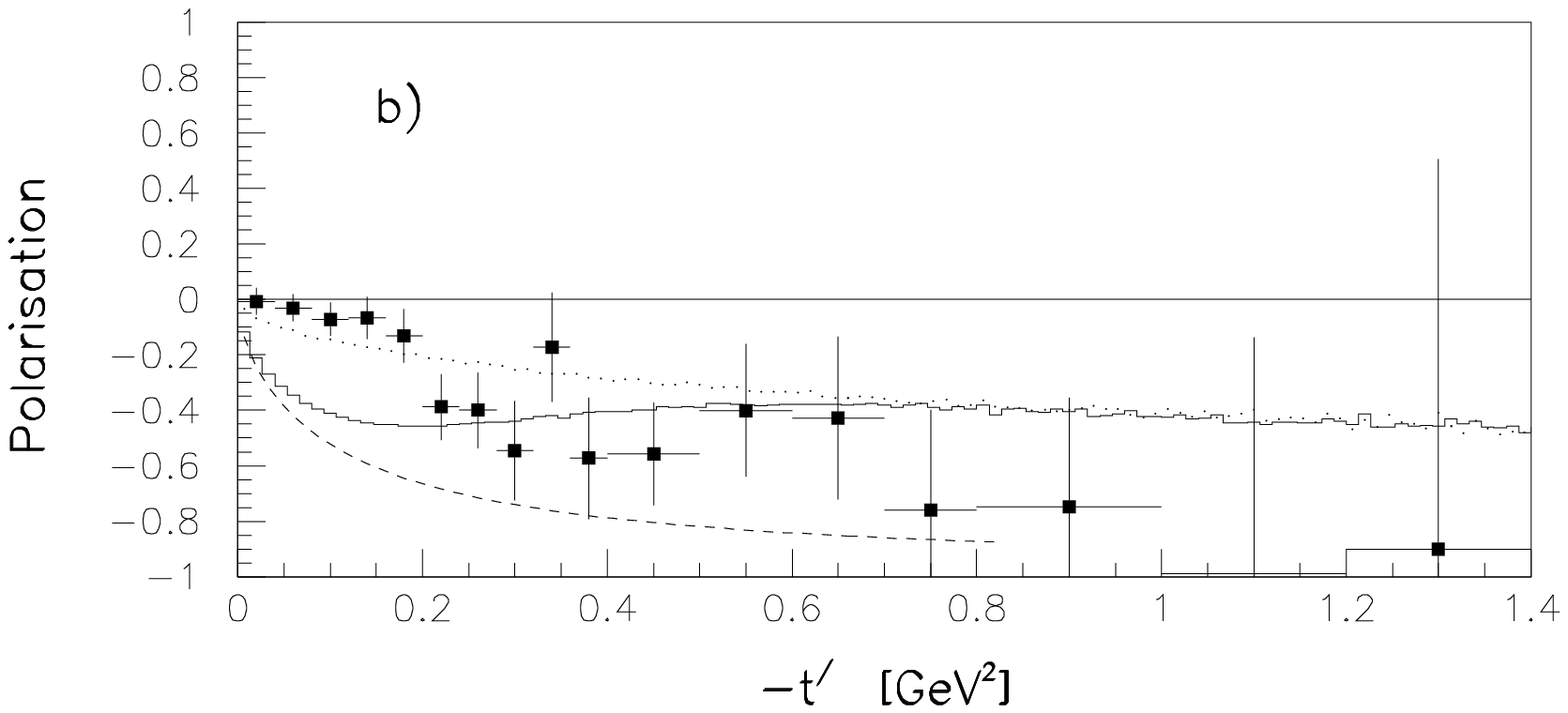}}
  \end{center}
  \caption{
    Experimental data \protect\cite{becker} on 
    (a) the differential cross section and (b) the polarisation
    in {\llb} taken at $p_{\overline{p}} = 6.0$~GeV/c 
    compared to string model simulations normalised to the total 
    cross section of the data.  
    The gaussian diquark $q_T$ distribution has 
    $\sigma_{q_T} = 0.45$~GeV (dotted line), 
    $\sigma_{q_T} = 0$~GeV (dashed line) and 
    a 40\%/60\% mixture of these two (full line).
    In all cases \mbox{$\sigma_{k_T} = 0.3$ GeV} is used.
 }
  \label{fig:becker}
\end{figure}

Fig.~\ref{fig:becker} shows experimental results for the
differential cross section for {\llb} obtained at 
$p_{\overline{p}} = 6.0$ GeV/c.
The data can be fitted with an empirical formula
with two exponentials
\begin{equation}
  d\sigma/dt' = a e^{bt'} + c e^{dt'}
  \label{eq:twoslope}
\end{equation}
with slope parameters 
$b = 10.1 \pm 0.6$~GeV$^{-2}$
and
$d =  3.0 \pm 0.3$~GeV$^{-2}$~\cite{becker}.
Here, 
  $t'=-\frac{1}{2}t_{max}'\left(1-\cosscat\right)$ 
is the reduced four--momentum transfer squared, where mass effects
are removed from the full four--momentum transfer 
\[  t = m^2_p + m^2_\Lambda - 2s + \frac{1}{2}t_{max}'\cosscat \]
with 
  $t_{max}'=\sqrt{(s-4m^2_p)\cdot(s-4m^2_\Lambda)}$. 

The shape of the differential cross section obtained with the string model is
also exponential in $t'$, 
since 
\mbox{$t' \simeq p_T^2$}
and, thereby, 
given by the gaussian $q_T$ and $k_T$ distributions. 
Using $\sigma_{q_T}=0.45$~GeV and $\sigma_{k_T}=0.3$~GeV 
one obtains the dotted line in Fig.~\ref{fig:becker}a.
These values are slightly lower than those 
(\mbox{$\sigma_{q_T}=0.5$~GeV} and \mbox{$\sigma_{k_T}=0.35$~GeV})
used in \cite{lund} to successfully describe {\lam} polarisation
in $pA \rightarrow \Lambda X$ at higher energies,
but lead to practically the same results in that domain.
The agreement of the model with the experimental $t'$-distribution 
is good for \mbox{$|t'| \gsim 0.4$~GeV$^2$}.
For lower values of $|t'|$ one obtains a good fit by setting 
$\sigma_{q_T} = 0$~GeV (dashed line). These two parameter settings give 
the slopes 
\mbox{$b = 10.1$~GeV$^{-2}$} and \mbox{$d =  3.0$~GeV$^{-2}$} 
in good agreement with the experimental values.
The Monte Carlo nature of the model makes it straight forward to add components
with different parameter settings. A sum of simulations using a 
60\% contribution with $\sigma_{q_T} = 0$~GeV and a 40\% contribution with
\mbox{$\sigma_{q_T} = 0.45$~GeV} results in the solid line in 
Fig.~\ref{fig:becker}a which reproduce the observed $t'$-distribution 
very well. 

We now turn to the polarisation which is shown as a function of $t'$ in 
Fig.~\ref{fig:becker}b. The original model with $\sigma_{q_T}=0.45$~GeV 
gives a decent description of the observed polarisation. 
With $\sigma_{q_T} = 0$~GeV, the only source for transverse momentum is 
given by the $s$ ($\bar{s}$) quark momentum resulting in a very strong 
correlation between $p_T$ and the polarisation. This is reflected in the 
rapid increase of the polarisation for small values of $|t'|$ shown by 
the dashed curve in Fig.~\ref{fig:becker}b. The full curve represents 
the above 60\%/40\% mixture and gives a reasonable description of the 
polarisation for $|t'| \gsim 0.2$~GeV$^{2}$, whereas at smaller $|t'|$ the 
contribution of \mbox{$\sigma_{q_T} = 0$} overestimates the magnitude 
of the polarisation.

High precision data on the {\llb} reaction closer to threshold have
been obtained by the PS185 collaboration at LEAR, CERN. 
A common feature in these data is that the angular distributions show, 
apart from the forward peaking, a more isotropic behaviour. 
This feature can not easily be explained by the model developed for 
higher energies. 
It can, however, 
be accommodated by adding an isotropic part to the diquark
angular distribution. 
This would lead to zero polarisation since 
each {\lam} scattering angle contains contributions
of $\vec{k}_T$ and $-\vec{k}_T$ with equal probability.

\begin{figure}[thp!]
  \begin{center}
  \resizebox{0.85\textwidth}{!}
  {\includegraphics*[1cm,5cm][20cm,21.5cm]{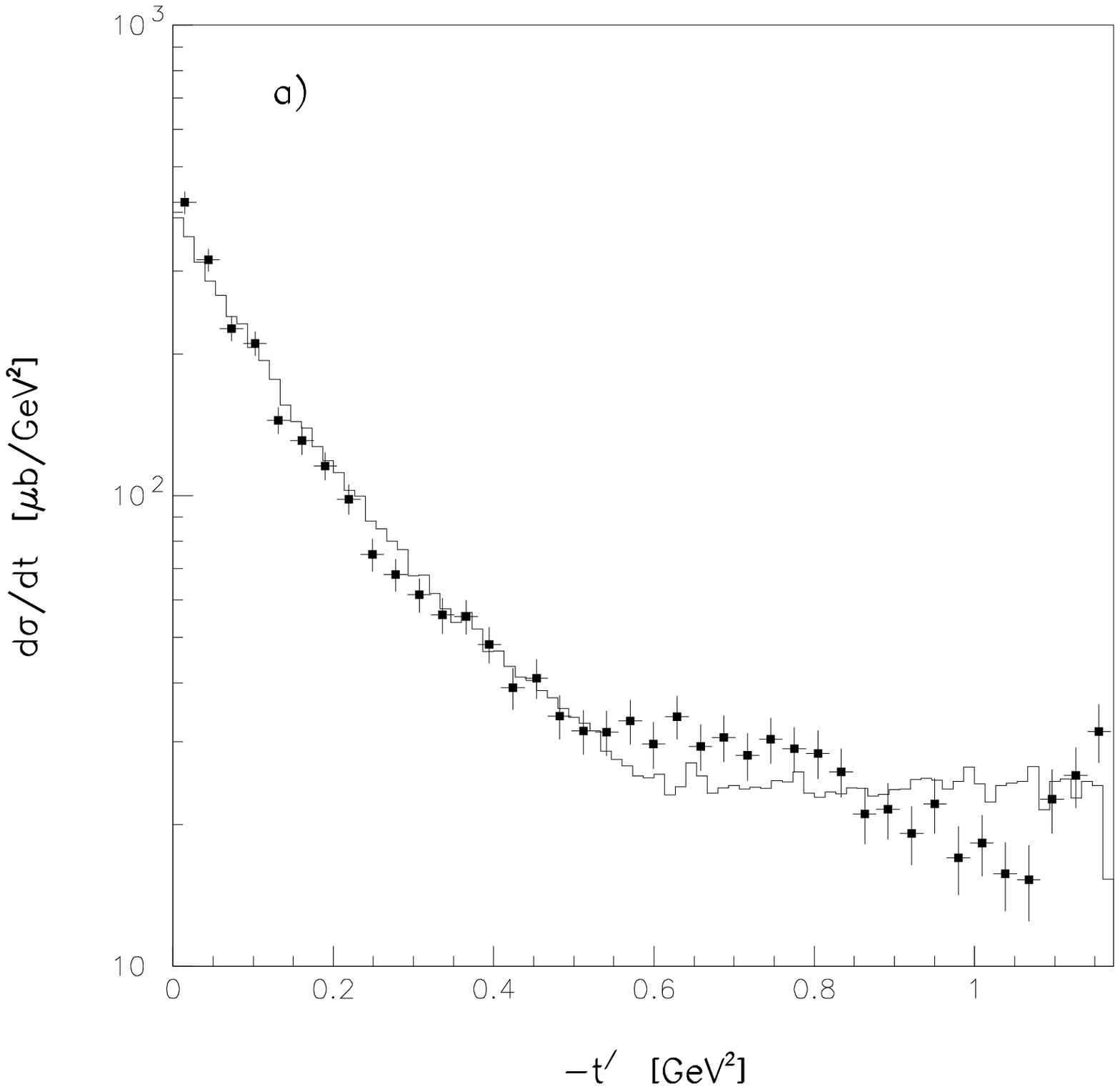}}
  \resizebox{0.85\textwidth}{!}
  {\includegraphics*[1cm,13cm][20cm,22cm]{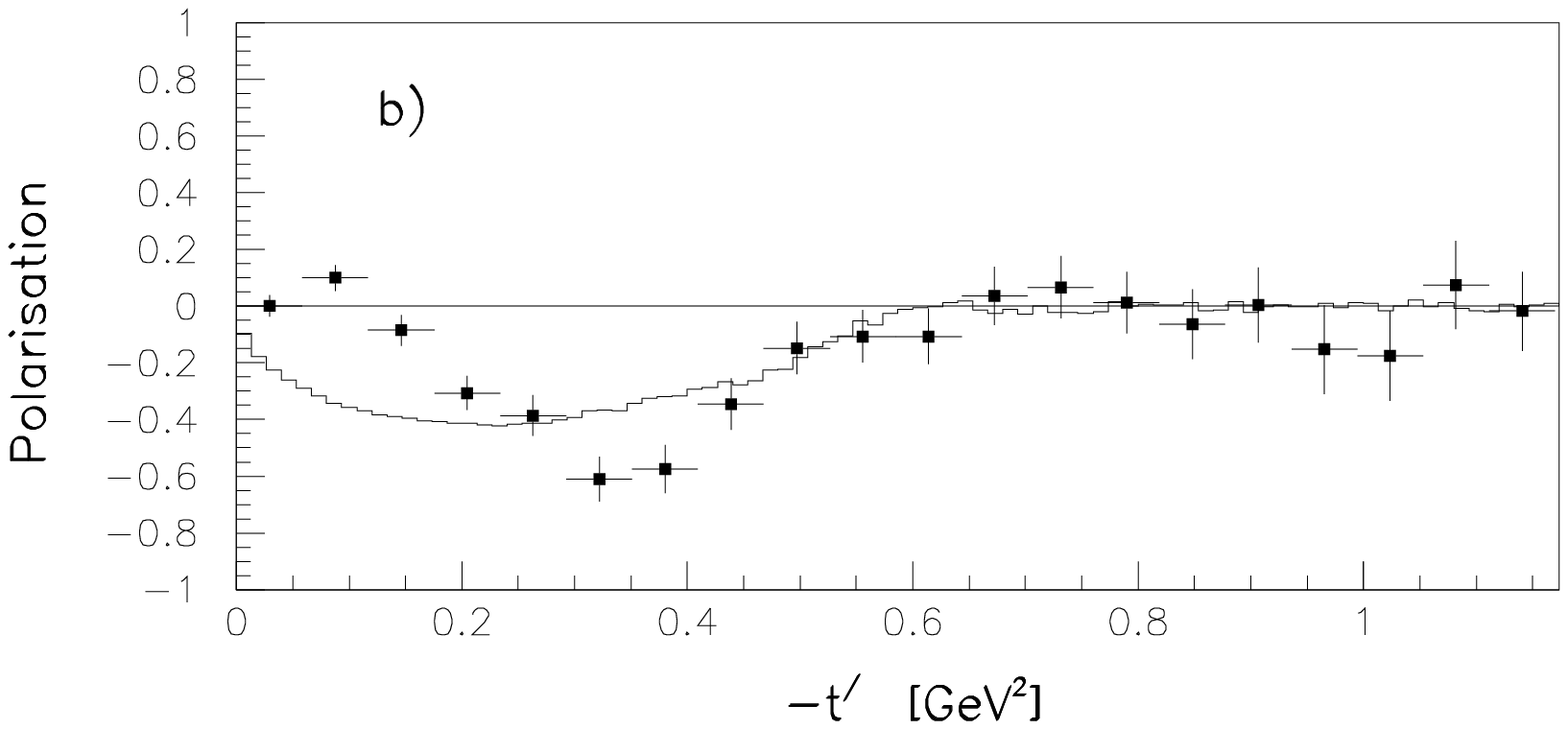}}
  \end{center}
  \caption{
    Comparison of the experimental data~\protect\cite{mac} at
    $p_{\overline{p}} = 1.835$~GeV/c with a string model simulation 
    having a 33\% contribution of isotropic diquark scattering and 
    the remainder with $\sigma_{q_T} = 0$~GeV.
  }
  \label{fig:lund_1835}
\end{figure}

Fig.~\ref{fig:lund_1835} shows the simulated differential cross section 
with an added isotropic diquark component, and polarisation for {\llb} at 
1.835 GeV/c.
No contribution with $\sigma_{q_T} = 0.45$~GeV is needed at this energy.
The comparison with the data from PS185
shows that the polarisation is also here
reasonably well reproduced for
\mbox{$|t'| \gsim 0.2$~GeV$^{2}$}.
Note that the polarisation becomes zero at $t'\approx -0.6$~GeV$^2$, 
corresponding to the c.m.s. scattering angle 
\mbox{$\Theta_{\overline{\Lambda}}^{*} = 90^\circ$}
in agreement with experimental results for
\mbox{$p_{\pbar} > 1.8$ GeV/c}
\cite{ps185_pol_data}.

\begin{figure}[thp!]
  \begin{center}
  \resizebox{0.85\textwidth}{!}
  {\includegraphics*[1cm,5cm][20cm,21.5cm]{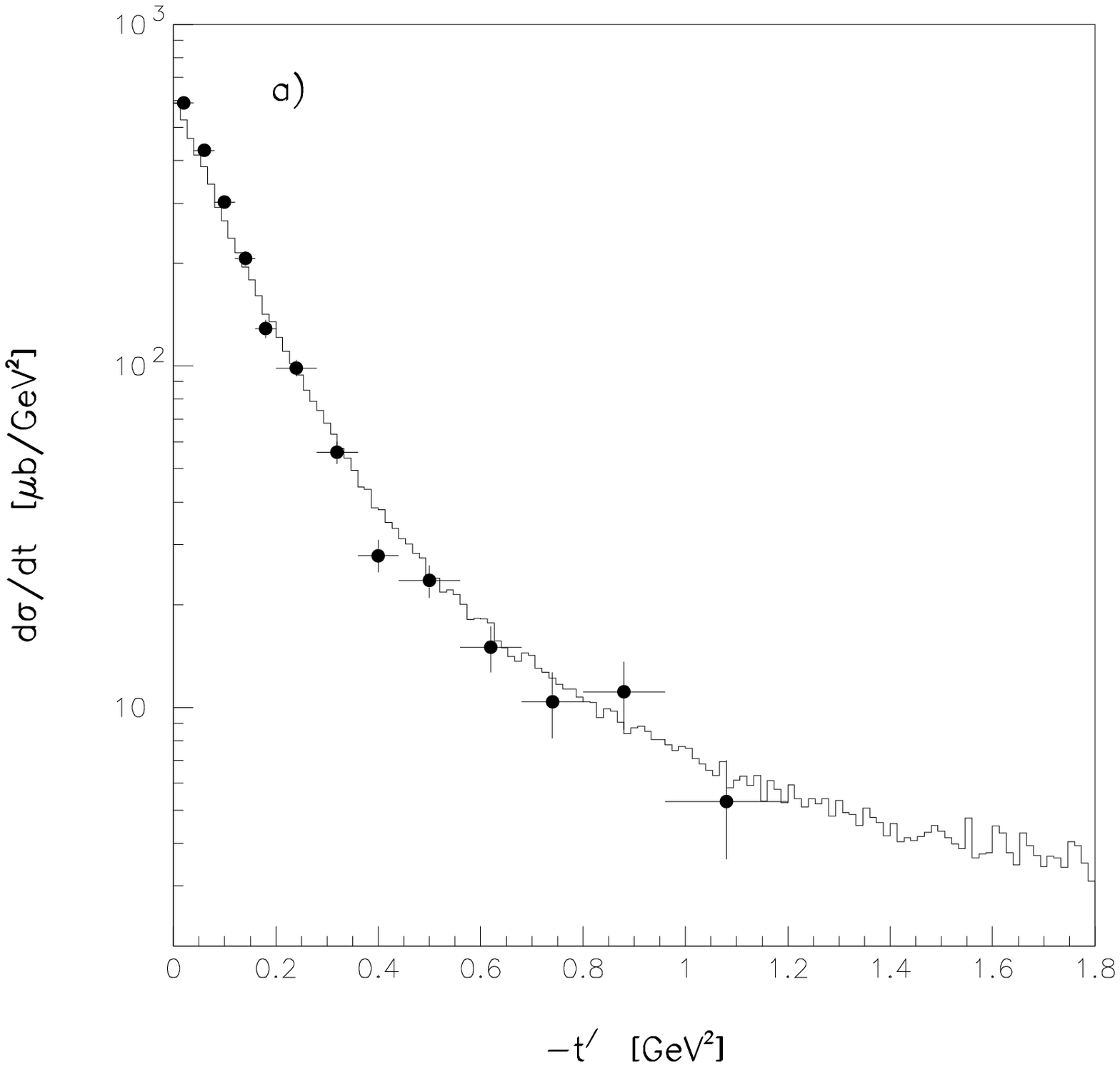}}
  \resizebox{0.85\textwidth}{!}
  {\includegraphics*[1cm,13cm][20cm,22cm]{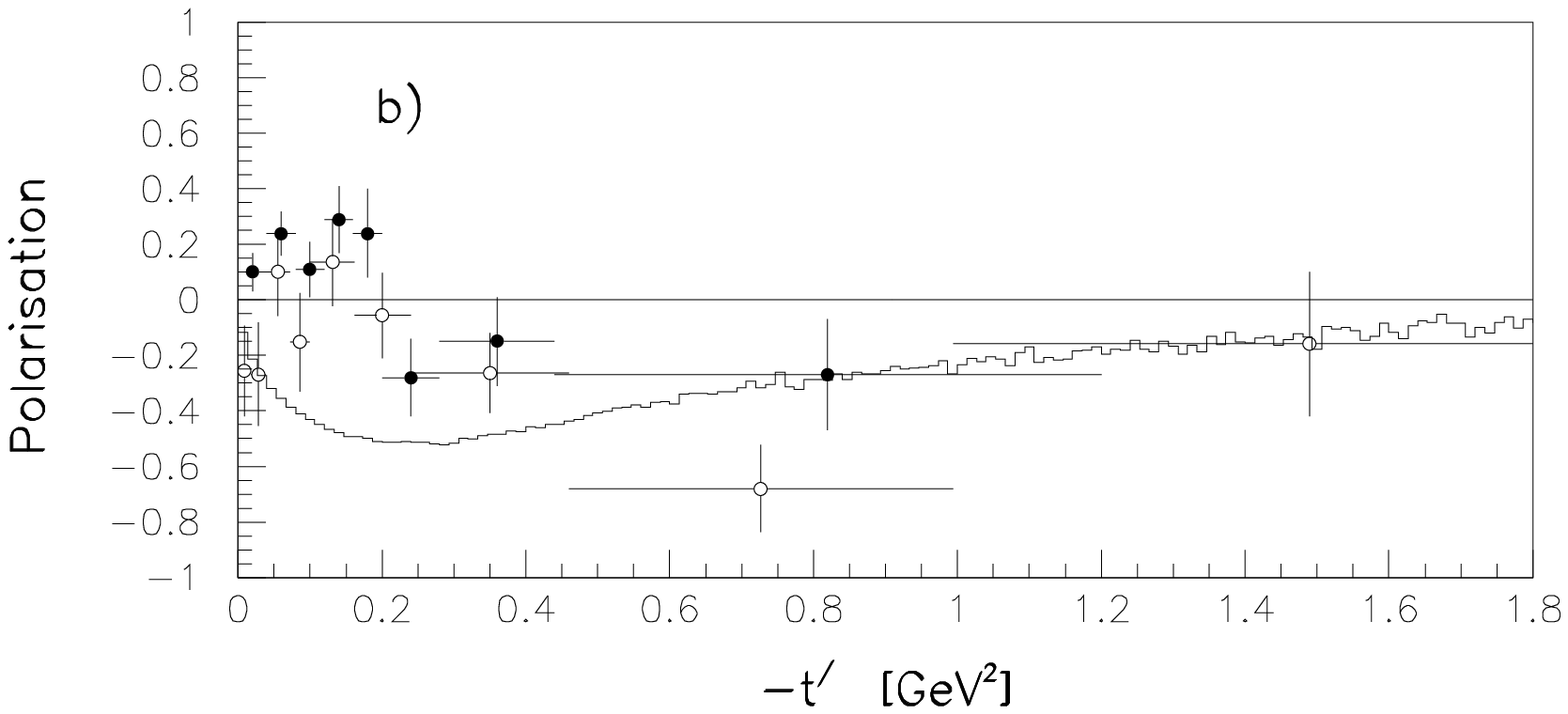}}
  \end{center}
  \caption{
   Comparison of experimental data 
   (solid circles for $p_{\overline{p}} = 3.0$~GeV/c \protect\cite{jacobs}
   and open circles for $p_{\overline{p}} = 3.6$~GeV/c \protect\cite{atherton})
   for
   (a) the differential cross section and
   (b) the polarisation
   with string model simulations at $p_{\overline{p}} = 3.0$~GeV/c
   (normalised to that data set). The diquark scattering has a 10\% 
   isotropic contribution, a 20\% contribution with $\sigma_{q_T} = 0.45$~GeV 
   and a 70\% contribution with $\sigma_{q_T} = 0$~GeV 
   (\mbox{$\sigma_{k_T} = 0.3$ GeV} always).
  }
  \label{fig:jacobs}
\end{figure}

Fig.~\ref{fig:jacobs} shows the result of a simulation made at
$p_{\overline{p}} = 3.0$~GeV/c 
together with data from Refs.~\cite{jacobs,atherton}.
For this intermediate energy, we again need 
a contribution with forward scattered diquarks ($\sigma_{q_T} = 0.45$~GeV).
The relative importance of the different contributions are discussed in 
next section. 
Table~\ref{table:param} gives an overview of the 
parameters of Eq.~\ref{eq:twoslope} obtained from the model and 
fitted to the experimental data.

\begin{table}[hbt!]
\begin{center}
\caption{
    Exponential slope parameters in Eq.~\protect(\ref{eq:twoslope}) 
    obtained by experiments in comparison with model simulation results.
}
\vspace{0.5cm}
\begin{tabular}{lllll}
\hline\noalign{\smallskip}
   {\pbar} momentum     &  \multicolumn{2}{c}{experiment} & 
   \multicolumn{2}{c}{string model} \\
  (GeV/c) 
             &  \multicolumn{1}{c}{$b$}
             &  \multicolumn{1}{c}{$d$}
             &  \multicolumn{1}{c}{$b$}
             &  \multicolumn{1}{c}{$d$}         \\
\noalign{\smallskip}\hline\noalign{\smallskip}
 6.0 \protect\cite{becker} & $10.1 \pm 0.6$ & $3.0 \pm 0.3$ &  10.1 & 3.0  \\
 3.6 \protect\cite{atherton}
                    & $8.4 \pm  0.4^\dagger$ & $2.0 \pm 0.2^\dagger$ &  9.0 & 1.8  \\
 3.0 \protect\cite{jacobs} & $9.66 \pm 0.47$ & $1.58 \pm 0.55$ & 8.9 & 1.8  \\
 1.8--2.0 \protect\cite{ralfandziol}
                           &    8--9         &      --      & 6.5--7.5 & -- \\
\noalign{\smallskip}\hline
\multicolumn{5}{l}{$^\dagger$These values are obtained in our parameterisation}\\
\multicolumn{5}{l}{although \protect\cite{atherton} use a different parameterisation.}
\end{tabular}
\label{table:param}
\end{center}
\end{table}
 
\section{Discussion and conclusions}
\label{sec:dis}

The Lund string model \cite{stringmodel} for the hadronisation of high energy 
quarks and gluons gives a natural explanation for the polarisation of 
$\Lambda$'s produced in the inclusive reaction $pA\rightarrow \Lambda X$
at high energy \cite{lund}. 
A $(ud)_0$ diquark is here assumed to be 
scattered with a transverse momentum $q_T$ and stretching 
a colour string-field. In the breaking of this field an $s$ quark
with transverse momentum $k_T$ is produced and a $\Lambda= (ud)_0s$
is formed with $\vec{p}_T= \vec{q}_T+\vec{k}_T$. 
The $s$ quark is polarised 
to compensate for an orbital angular momentum proportional to $k_T$, 
giving a polarisation of the $\Lambda$ which increases with its $p_T$.
With the normal gaussian distributions in $k_T$ and $q_T$ used in 
the Lund model, 
the observed $\Lambda$ polarisation is reproduced \cite{lund}. 

An extension of this model has been developed here in order to test whether 
the same polarisation mechanism is able to reproduce also the 
polarisation observed in the two-body {\llb} reaction at low energy. 
The $s\bar{s}$ pair production in the string-field is supposed to be a 
local phenomenon in the string and is therefore expected to be energy 
independent. 
Therefore, we have not altered the gaussian $k_T$ distribution 
for the strange quarks, which is also the essential source of the polarisation. 
The diquark scattering mechanism, on the other hand, may very well be 
energy dependent. We have therefore extended the diquark scattering model 
to include three components: 
(1) forward scattered diquarks with \mbox{$\sigma_{q_T} = 0.45$ GeV},
(2) diquarks as spectators without transverse momenta and
(3) isotropic diquark scattering.

The first component, which was the only one included in the original model 
applied at high energies, 
becomes less important with decreasing $\overline{p}$ momentum;
40\% at 6 GeV/c, 20\% at 3 GeV/c and neglible below 2.0~GeV/c.
On the other hand, the isotropic part is becoming more important closer
to threshold.
It is negligible at 6 GeV/c but increases to about 30\% below 2 GeV/c.
This reflects the fact that the 
$\overline{\Lambda}\Lambda$ pairs are produced in an S-wave at threshold. 
Diquarks as spectators contribute 60--70\% at all energies.
Since here, the diquarks have no transverse momenta (\mbox{$\sigma_{q_T} = 0$ GeV}),
it is only the ${k_T}$ of the strange quark that provides the transverse momentum 
and, therefore, \mbox{$\sigma_{k_T}$} gives the slope of the $t'$ distribution. 
This is reflected in the sharp forward rise 
in the angular distribution which is typical behaviour for peripheral processes.
Using a the black--disc model, a slope of \mbox{8~GeV$^{-2}$} would
correspond to an absorption radius of about 1~fm.

This string--based model cannot explain positive polarisation as observed in 
the region \mbox{$|t'| < 0.2$ GeV$^{2}$}. 
In the semiclassical picture we apply, this means that the
diquarks pass each other at a distance of the order of
the absorption radius, 
leading to an orbital angular momentum in the 
($ud$)--($\overline{u}\overline{d}$) system.
One may, therefore, imagine a scenario where both
the orbital angular momentum of the $\overline{s}s$--pair
and their spins compensate the angular momentum of the
($ud$)--($\overline{u}\overline{d}$) system.
This would lead to positive polarisation in the very forward region
(small $k_T$).
For larger $k_T$ the $s\overline{s}$ angular momentum
will overcompensate the ($ud$)--($\overline{u}\overline{d}$)
angular momentum and we would again expect negative polarisation.
To quantify this picture is, however, complicated since the 
orientation of the string, and thus $\vec{k}_T$ at the breakup, 
will vary with time. 

The fact that the string model works quite well for larger momentum transfers,
but not for smaller may give more fundamental insights. The border line 
\mbox{$|t'| \approx 0.2$ GeV$^{2}$} corresponds to a momentum transfer of 
a few hundred MeV. 
Above that, it is reasonable that a model formulated 
in a quark basis should be applicable since quark degrees of freedom can be 
resolved. For lower momentum transfers, however, even a constituent quark 
structure may be unresolved and it may be more appropriate to have a 
description in terms of hadrons. Such models formulated in a hadron basis 
have been considered in this context of {\llb}, 
see, {\it e.g.}, \cite{hadronmodel} and references therein. 
With suitable components of $K$ and $K^*$ exchange,
including initial--state and final--state interactions,
it is possible to obtain both the positive and negative polarisation. 
At this low energy scale it is not surprising that 
the full dynamical behaviour cannot be well described within the same simple
quark--based model that works well at high energies. 

In view of this, it is quite remarkable 
that the string--based model with only few parameters
works so well for {\llb} at low energies. 
Both the observed angular distribution and, 
for not too small momentum transfers, 
also the polarisation in the $\overline{\Lambda}\Lambda$ system 
are well described. 
Moreover, the model gives a natural explanation for the fact
that $\overline{\Lambda}\Lambda$ are produced in a spin triplet state.
In total, this gives evidence for a universal origin of the $\Lambda$ 
polarisation phenomena observed at different energies.


\bibliographystyle{unsrt_notitle}
\bibliography{lund_ref_new}

\end{document}